\documentstyle[psfig]{mn}

\newif\ifAMStwofonts

\ifoldfss
  \ifCUPmtlplainloaded \else
    \NewTextAlphabet{textbfit} {cmbxti10} {}
    \NewTextAlphabet{textbfss} {cmssbx10} {}
    \NewMathAlphabet{mathbfit} {cmbxti10} {} 
    \NewMathAlphabet{mathbfss} {cmssbx10} {} 
  \fi
  \ifAMStwofonts
    \ifCUPmtlplainloaded \else
      \NewSymbolFont{upmath} {eurm10}
      \NewSymbolFont{AMSa} {msam10}
      \NewMathSymbol{\upi}     {0}{upmath}{19}
      \NewMathSymbol{\umu}     {0}{upmath}{16}
      \NewMathSymbol{\upartial}{0}{upmath}{40}
      \NewMathSymbol{\leqslant}{3}{AMSa}{36}
      \NewMathSymbol{\geqslant}{3}{AMSa}{3E}

      \let\leq=\leqslant 
       \let\ge=\geqslant
    \fi
  \fi
\fi 

\ifnfssone
  \newmathalphabet{\mathit}
  \addtoversion{normal}{\mathit}{cmr}{m}{it}
  \addtoversion{bold}{\mathit}{cmr}{bx}{it}
  \newmathalphabet{\mathbfit} 
  \addtoversion{normal}{\mathbfit}{cmr}{bx}{it}
  \addtoversion{bold}{\mathbfit}{cmr}{bx}{it}
  \newmathalphabet{\mathbfss} 
  \addtoversion{normal}{\mathbfss}{cmss}{bx}{n}
  \addtoversion{bold}{\mathbfss}{cmss}{bx}{n}
  \ifAMStwofonts
    \ifCUPmtlplainloaded \else
      \UseAMStwoboldmath
      \makeatletter
      \new@mathgroup\upmath@group
      \define@mathgroup\mv@normal\upmath@group{eur}{m}{n}
      \define@mathgroup\mv@bold\upmath@group{eur}{b}{n}
      \edef\UPM{\hexnumber\upmath@group}
      \new@mathgroup\amsa@group
      \define@mathgroup\mv@normal\amsa@group{msa}{m}{n}
      \define@mathgroup\mv@bold\amsa@group{msa}{m}{n}
      \edef\AMSa{\hexnumber\amsa@group}
      \makeatother
      \mathchardef\upi="0\UPM19
      \mathchardef\umu="0\UPM16
      \mathchardef\upartial="0\UPM40
      \mathchardef\leqslant="3\AMSa36
      \mathchardef\geqslant="3\AMSa3E

      \let\leq=\leqslant 
       \let\ge=\geqslant
    \fi
  \fi
\fi 

\ifnfsstwo
  \DeclareMathAlphabet{\mathbfit}{OT1}{cmr}{bx}{it}
  \SetMathAlphabet\mathbfit{bold}{OT1}{cmr}{bx}{it}
  \DeclareMathAlphabet{\mathbfss}{OT1}{cmss}{bx}{n}
  \SetMathAlphabet\mathbfss{bold}{OT1}{cmss}{bx}{n}
  \ifAMStwofonts
    \ifCUPmtlplainloaded \else
      \DeclareSymbolFont{UPM}{U}{eur}{m}{n}
      \SetSymbolFont{UPM}{bold}{U}{eur}{b}{n}
      \DeclareSymbolFont{AMSa}{U}{msa}{m}{n}
      \DeclareMathSymbol{\upi}{0}{UPM}{"19}
      \DeclareMathSymbol{\umu}{0}{UPM}{"16}
      \DeclareMathSymbol{\upartial}{0}{UPM}{"40}
      \DeclareMathSymbol{\leqslant}{3}{AMSa}{"36}
      \DeclareMathSymbol{\geqslant}{3}{AMSa}{"3E}

      \let\leq=\leqslant 
       \let\ge=\geqslant
    \fi
  \fi
\fi 

\ifCUPmtlplainloaded \else
  \ifAMStwofonts \else 
    \def\upi{\pi}
    \def\umu{\mu}
    \def\upartial{\partial}
  \fi
\fi

\title[XTE J0052-723]{X-Ray and Optical Observations of XTE J0052-723 a Transient 
Be/X-Ray Pulsar in the Small Magellanic Cloud}
\author[Laycock, Corbet, Coe, Marshall, Markwardt, Edge]
       {S.~Laycock$^1$, R. H. D.~Corbet$^{2,3}$, M. J.~Coe$^1$, F. E.~Marshall$^2$, C.~Markwardt$^{2,4}$, W.~Edge$^1$  \\
       $^1$Department of Physics and Astronomy, Southampton University, SO17 1BJ \\
       $^2$NASA Goddard Space Flight Center, Greenbelt, MD 20771 USA \\
       $^3$Universities Space Research Association \\
       $^4$Department of Astronomy, University of Maryland, College Park, MD 20742, USA}
\date{Accepted .
      Received ;
      in original form 
      }

\pagerange{\pageref{firstpage}--\pageref{lastpage}}
\pubyear{2001}

\begin{document}

\maketitle

\label{firstpage}

\begin{abstract}
On December 27th 2000 during our regular SMC monitoring program with Rossi X-ray 
Timing Explorer ($RXTE$) strong pulsations were detected with a period 
of 4.78 seconds. Subsequent slew observations performed on Jan 9$^{th}$ and 13$^{th}$  
across the field of view allowed localisation of the pulsar's position 
to RA: 0$^h$52$^m$17$^s$, Dec: 72$\degr$19$\arcmin$51$\arcsec$ (J2000).  
The outburst continued until Jan 24$^{th}$, 7 PCA observations were obtained during 
this period, yielding a maximum X-ray luminosity $\sim$10$^{38}$ ergs/s. 
Following calculation of the pulsar position, optical observations of the $RXTE$ 
error box were made on Jan 16$^{th}$ 2001 with the 1m telescope of the South African 
Astronomical Observatory (SAAO) while the source was still in X-ray 
outburst. Candidate Be stars identified from their photometric colours were 
subsequently observed with the SAAO 1.9m on Nov 7$^{th}$ 2001 to obtain spectra. 
Only one of the photometrically identified stars [MA93]537 showed 
prominent H$\alpha$ emission, with a double peaked line-profile 
(EW= -43.3$\pm$0.7\AA, separation velocity=200$\pm$15 km/s) 
confirming the presence of a substantial circumstellar disk. 
  
\end{abstract}


\section[]{Introduction}
High Mass X-ray Binaries (HMXBs) are traditionally divided on the basis of the nature of 
the optical counterpart, into Be and supergiant binary systems. A recent survey of the 
literature reveals that of the 96 proposed massive X-ray binary pulsar systems, 67\% of the
identified systems fall within the Be/X-ray group of binaries \cite{coe00}. 
In the Small Magellanic Cloud (SMC), almost all of the systems identified so far 
(except SMC X-1, a B0I Supergiant) are believed to be of this type \cite{hs00}. 
The orbit of the Be star and the compact object, a neutron star, is generally wide and 
eccentric. The optical star exhibits H$\alpha$ line emission and continuum 
free-free emission (revealed as excess flux in the IR) from a disk of 
circumstellar gas. The Be star disk phenomenon is a consequence of these stars' 
rapid rotation, which is typically $>$60\% of breakup velocity \cite{porter96} 
Most of the Be/X-ray sources are transient in the emission of X-rays due 
to a combination of orbital motion and timescales for changes in the Be star disk. 
X-ray outbursts appear to fall into two classes on the basis of luminosity 
and periodicity: So-called 'normal' outbursts reach $\sim$10$^{37}$ ergs/s 
and are associated with periastron passage of the neutron star through the 
Be star's circumstellar disk. Giant outbursts may reach over an order of 
magnitude brighter and occur sporadically, uncorrelated with the orbital 
phase of the system. See for example Negueruela \shortcite{neg98} for a review.

The X-ray pulsar population of the SMC is far greater than expected on the basis of the
its size. Although only $\sim$1/50 the mass of the Milky Way, the SMC contains at least 1/3 as many
HMXBs, i.e. 24 known SMC pulsars rather than the expected $\sim$1. Given the expected
lifetime of a Be/X-ray binary is $<$1My \cite{vdh83}, the existence of so many implies they all formed 
at about the same
time. In the standard evolutionary scenario of van den Heuvel \shortcite{vdh83} 15 My elapses between 
formation of a massive binary and its transition to an HMXB. The exact nature of the event which triggered
star formation in the SMC 15-20 My ago is unclear, however there are important clues from dynamical 
simulations and recent photometric \& radio surveys.
Simulations of tidal interactions between the Magellanic clouds and Milky Way over the past 2 Gy 
by Gardiner \& Noguchi \shortcite{GN96} indicate that the Clouds have experienced two  
close encounters, 1.5 \& 0.2 Gy ago. During the most recent interaction, the clouds passed 
within 14 kpc of each other and the total tidal force exerted on the SMC rose by a factor of $\sim$30. 
Subsequent extension of this model to include star formation (Yoshizawa \& Noguchi 1999, Gardiner
1999) appears to predict a time-lag between the 
tidal disruption and resulting star formation rate, which may only be peaking today. Observational 
evidence shows that the SMC consists of two distinct stellar populations \cite{Z00}. 
The older (pre LMC encounter) population shows a smooth, spheroidal distribution. In contrast, 
the main sequence population is highly irregular, with the very youngest stars concentrated almost 
entirely in the 'Bar' and 'Wing' regions \cite{M01}. These are also the regions in which the majority 
of X-ray pulsars have been found. There are still inconsistencies in this overall picture which point 
to the influence of hydrodynamic processes \cite{LT99}, indeed the complex structure of the ISM shows 
clear evidence of having been sculpted by supernovae and winds from massive stars. The radio survey of 
Stanimirovic \shortcite{S99} reveals that the large-scale structure of giant HI shells in the SMC shows
evidence of hierarchical arrangement. Smaller shells are distributed around the larger ones, suggesting that 
star formation has proceeded in an impulsively triggered sequence, rather than simultaneously due to 
large-scale gravitational collapse.     

A detailed investigation of the luminosity and period distributions and long-term behaviour of 
X-ray pulsar systems in the SMC is underway. The low extinction and unobstructed
line of sight to the SMC enable accurate measurements of X-ray luminosity and useful 
observations of the optical counterparts with 1-2m telescopes. Since the distance to 
the SMC is significantly greater than its 'depth' all of the objects 
in the SMC can be considered to lie at an equal distance for the purposes of luminosity 
calculations. 

The source that is the subject of this paper, XTE J0052-723
\cite{corbet2001} is thought to be a new Be/X-ray binary system. It was detected in 
the course of our regular pulsar search program of the SMC, which has been running 
since January 1999, previously described by Laycock et al. \shortcite{lay01}. At the time of
XTE J0052-723's discovery our observing strategy employed approximately weekly observations 
with a duration of $\sim$5 ksec centred on RA: 0$^h$50$^m$ DEC: -73$\degr$6$\arcmin$, in the
south eastern arm of the SMC. From this position the PCA's large 2\degr wide field of view 
encompasses the positions of 16 of the 24 confirmed X-ray pulsars in the SMC. The range of 
known pulse periods spans 0.716 - 755 seconds. 
  
\section{X-ray Observations}
\subsection{RXTE PCA}
The {\it RXTE} Proportional Counter Array (PCA) \cite{jahoda96} consists of 5 collimated 
proportional counter units (PCU), each with a collecting area of 1300 cm$^{2}$. 
The field of view of the PCA is approximately 2\degr FWZI with a triangular collimator 
response as a function of off-axis viewing angle. Each PCU has 3 xenon filled detector-anode 
layers, and an additional upper layer filled with propane. The propane layer is operated as 
a veto detector in order to reduce particle background.  
In certain data modes (e.g. Good Xenon and Standard 2) separation of data by detector  
is maintained in the telemetry, allowing data from each layer and PCU to be analysed 
independently. The highest timing and spectral resolution are obtained in Good Xenon mode, 
in which the energy and arrival time of of every detector event is registered. The timing resolution 
is 1$\mu$s with an absolute accuracy limited by the spacecraft clock to 8$\mu$s, photon energy 
is recorded by a 256 channel pulse-height discriminator with a resolution of 18\% at 6 keV. Standard 2
mode provides time-binned data with 16s resolution and 129 energy channels. Both of these data 
modes were used in our analysis. The signal-to-noise for low energy X-rays 2$\sim$25 keV 
is at its maximum in the uppermost xenon layer (layer 1), while the full (nominal) 2-60 keV 
energy range of the PCA is achieved when data from all 3 layers are combined.

\subsection{Monitoring and Detection} 
\label{section:detection}
In our SMC monitoring program we adopt an automated reduction/analysis
pipeline that is optimised for pulsar detection.
Good Xenon event mode data from the PCA \cite{jahoda96} are used to construct
lightcurves at 10ms timing resolution. In order to maximise signal to
noise, only events from the top anode layer of 
the PCA are included. We use the energy range 3-10 keV in order to cover the 
peak emission expected for X-ray pulsars. Few filtering criteria are applied, 
in order to preserve our time coverage and hence temporal resolution. 
Background count rates are estimated using the L7 PCA faint background models.  
The count-rates are then normalised to Counts PCU$^{-1}$ s$^{-1}$ and bin times 
corrected to the solar-system barycentre. 
 
For the detailed study of the new pulsar, additional reduction steps were
required to eliminate systematic uncertainties introduced by the defective PCU 0. 
Since the loss in May 2000 of the propane layer on PCU 0, the background event 
rate in this detector no longer matches the L7 background model. Additionally we 
found significant differences between count rates and spectra obtained from PCU 0 and 
the rest of the PCA and consequently excluded PCU 0 data from our analysis. 
The active PCUs in each observation are indicated in Table~\ref{tab:obs}, taking 
$\surd(N_{PCU}T_{exposure})$, as indicative of the sensitivity of each observation, 
the range of relative sensitivity spans a factor $\sim$1.7 between observations 4 and 7. 
Data obtained within 30 minutes following passage of $RXTE$ through the South Atlantic 
Anomaly were excluded, as were data taken during any time the particle 
background exceeded 0.1 electrons/sec in PCU 2. These conditions were applied 
retrospectively to all observations within 3 weeks either side of the
outburst of XTE J0052-723 in order to obtain an accurate mean count rate with 
which to estimate the contribution from interfering sources in the field of view. 
\begin{table}
 \caption{Dates of Observations }
 \label{tab:obs}
 \begin{tabular}{lccl} 
 Obs.   &       Date        &  Observation       & Active PCUs	\\
\hline
 1      &       27/12/00    &  3.85ksec stare    & 0124       	\\
 2      &       03/01/01    &  5.12ksec stare    & 023		\\
 3      &       09/01/01    &  slews             & 0234		\\
 4      &       09/01/01    &  2.91ksec stare    & 0234 	\\
 5      &       13/01/01    &  slews             & 01234 	\\
 6      &       18/01/01    &  4.15ksec stare    & 01234	\\
 7      &       24/01/01    &  5.12ksec stare    & 01234	\\
\end{tabular}
\end{table}
\subsection{Source Location}
The position of the new pulsar was determined by slew observations on Jan 9th and 
13th 2001 (see Table~\ref{tab:obs}). Orthogonal slews were made in RA and DEC over 
slightly different position on each occasion, with the second set of slews used to 
further refine the source position determined from the first. As the source moves 
through the collimator field of view, it samples different values of the collimator 
response function, which modulates the detected count rate. Given a trial position 
and intensity for the source, a model light curve can be constructed using the 
response function, and compared to the observed light curve. The
position and intensity are then varied iteratively using least squares
fitting until the best-fit parameters are obtained. The 99\% confidence contour
determined for XTE J0052-723 is a rectangular region $RA: 0^h52^m17^s \pm24^s, 
Dec: -72\degr19\arcmin51\arcsec \pm50\arcsec$ (J2000).  

\subsection{Timing Analysis}
\label{section:timing}
The 10ms resolution lightcurves obtained from the 5 pointed observations were 
analysed using the Lomb-Scargle periodogram \cite{sca82} at a resolution of 
0.0001Hz (2.5ms or 0.05\% at 5s). The Lomb-Scargle method was used to calculate 
the power density spectrum (PDS) owing to gaps in the data caused by SAA passage, 
earth occultation and slews. Additionally this procedure is suited to  
the uneven time sampling introduced by barycentre correction of the lightcurve 
bin-times. The fast Lomb-Scargle code of Press \& Rybicki 
\shortcite{press89} was implemented in double precision. 

Owing to the highly non-sinusoidal nature of the pulse profiles, the periods 
found in the PDS were investigated using Phase Dispersion Minimization (PDM) 
\cite{stel78}. The lightcurves were folded on a series of trial periods centred 
on the period of interest, at each trial period the folded data were binned in 
10 (non overlapping) phase intervals to create a template. The variance of points 
belonging to each bin was then combined with the overall variance in the
lightcurve to create a measure of the goodness-of-fit between the derived 
template and the lightcurve. This approach has an advantage over Fourier techniques 
that use an assumed fitting function. We found that PDM detected the 4.782s fundamental 
pulse period as the strongest periodogram feature even in observations where the 
PDS showed only the P/2 harmonic. Resulting pulse periods and their errors
estimated from the PDM are given in Table~\ref{tab:periods}. In 
Figure~\ref{fig:pdm} the 4.78s feature is displayed for each observation in 
order to assess the evidence for changes in the pulse period. For this plot we 
show the PDM-statistic calculated for trial periods in the range 4.77 to – 4.79s 
which is appropriate for the width of the feature. 
\begin{table}
 \caption{Observed parameters of XTE J0052-723}
 \label{tab:periods}
 \begin{tabular}{llll}
 Obs.        & Pulse period          &  Pulse amplitude        & Mean flux            \\
 number      & seconds               &  Ct PCU$^{-1}$s$^{-1}$  & Ct PCU$^{-1}$s$^{-1}$\\
\hline
   1         &  4.7817  $\pm$ 0.0001 &  2.58 $\pm$0.71         & 11.4                \\
   2         &  4.7818 $\pm$ 0.0001  &  3.83 $\pm$0.69         & 11.8                 \\
   4         &  4.782  $\pm$0.001    &   3.88 $\pm$0.79        & 8.6		      \\	
   6         &  4.7820 $\pm$0.0001   &   2.88 $\pm$0.42        & 7.5		      \\	
   7         &  4.7823  $\pm$0.0001  &   2.04 $\pm$0.51        & 5.6                 \\
\end{tabular}

\medskip
Values are corrected for collimator response and interfering source contribution 
($\sim$0.9Ct PCU$^{-1}$s$^{-1}$). Appropriate collimator response is 0.21 for 
obs 1,2,4,7 \& 0.99 for obs 6. See Section~\ref{section:timing} for explanation.  
note: timing analysis not applied to slew observations 3 \& 5 
\end{table}

\begin{figure}
\begin{center}
\psfig{file=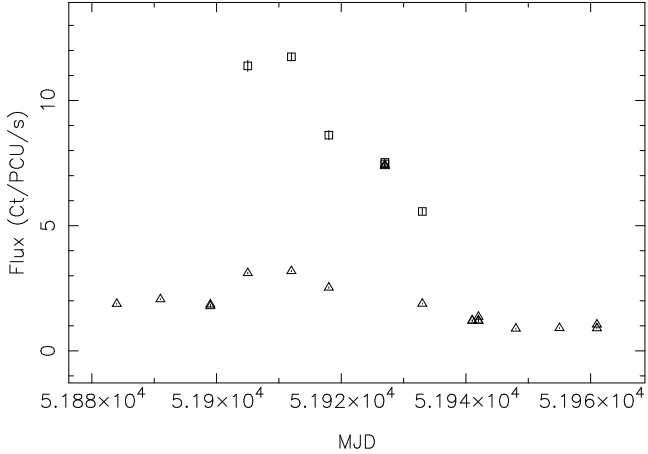,width=8.5cm,angle=-90}
\caption{Background corrected 3-10 keV count rate before, during, and after the outburst.
 Observations containing 4.78s pulsations have been corrected for the 
 collimator response to XTE J0052-723, and an interfering source component
 consistent with the last 3 flux values shown. $\bigtriangleup$= Observed flux, $\sq$= Calculated XTE J0052-723 
 flux with error bars.}
\label{fig:fluxes}
\end{center}   
\end{figure}

\begin{figure}
\begin{center}
\psfig{file=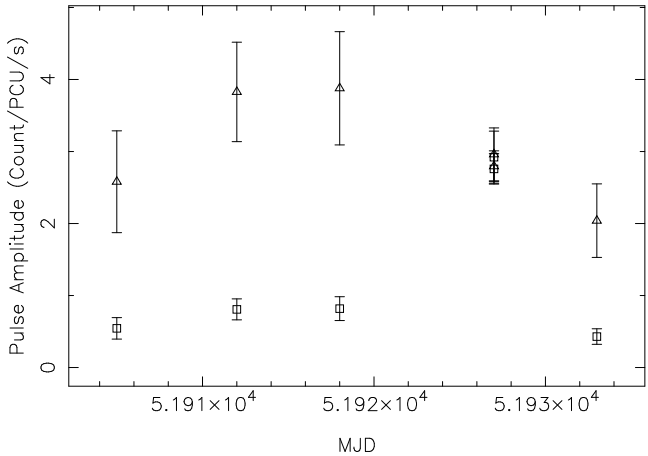,width=8.5cm,angle=-90}
\caption{Pulse amplitude during the outburst. $\bigtriangleup$= collimator corrected, $\sq$= raw data }
\label{fig:pulsetime}
\end{center}
\end{figure}

Having established the pulse period in each observation, the lightcurves were 
folded to produce pulse profiles. From these profiles peak-peak pulse amplitudes 
were obtained in units of Counts PCU$^{-1}$s$^{-1}$
Following determination of the source position (Section 2.1) the epoch-folded 
fluxes from each observation were corrected for collimator response due to their 
different pointing positions. 
\begin{figure} 
\begin{center}
\psfig{file=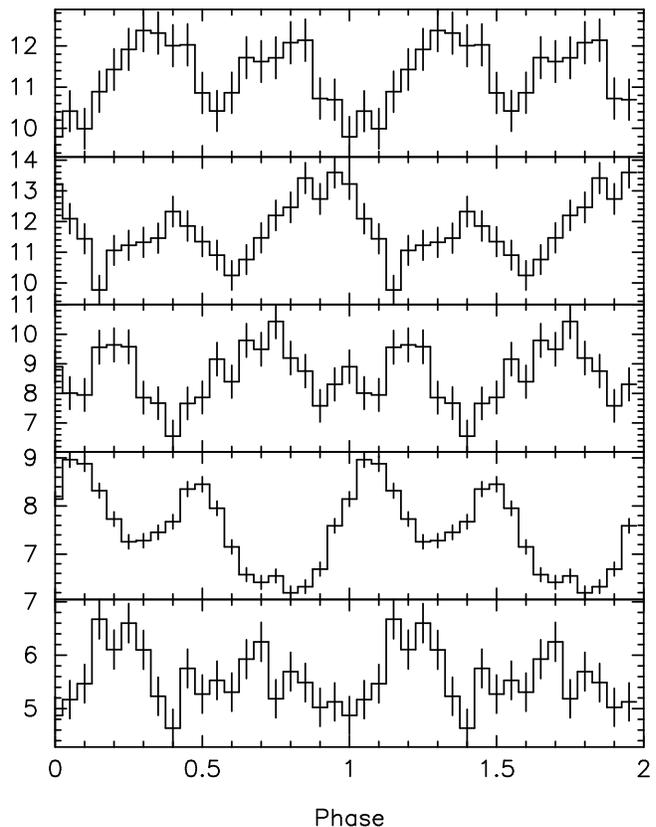,width=8.5cm}
\caption{Pulse profiles 3-10 keV, Counts PCU$^{-1}$s$^{-1}$, background 
subtracted, collimator corrected assuming a constant 0.89 Counts 
PCU$^{-1}$s$^{-1}$ interfering source component. Error bars have been increased in
proportion to the collimator correction. Arbitrary T$_{0}$=MJD51905. From top down, 
observations 1,2,4,6,7}
\label{fig:profiles}
\end{center}
\end{figure}
Although XTE J0052-723 was the only pulsar detected in these observations, there 
is expected to be a non zero contribution to the X-ray flux due to other sources 
in the SMC. No pointings were made at the exact co-ordinates of the new pulsar after 
it had faded from view and so it is difficult to estimate this component. 
However observations were made at the standard monitoring position (see Section 
2) before, during, and after the outburst. Our estimated value for the persistent 
background at this pointing was 0.9 Counts PCU$^{-1}$s$^{-1}$, which is
consistent with the post outburst flux seen for 3 consecutive weeks after the 
end of the observed outburst. Observations immediately before the outburst show 
higher fluxes and pulsations at $\sim$46s and $\sim$173s
which we attribute to the known X-ray pulsars 1WGA J0053.8-7226 \cite{singh95} 
and AX J0051.6-7311 \cite{yok00}. Figure~\ref{fig:fluxes} shows the mean count 
rate as a function of time during this period. For all observations showing 
4.78s pulsations the count rate has been corrected for the collimator response 
to the best-fit position and the 0.9 Counts PCU$^{-1}$s$^{-1}$ background contribution.  
As an additional indicator of the outburst profile, the variation in pulsed-flux 
was investigated as it is independent of assumptions about the background. 
Figure~\ref{fig:pulsetime} shows the pulse amplitudes given in Table~\ref{tab:periods} 
before and after correction for pointing offset (collimator response). The errors in 
these points are dominated by the propagation of errors in measured amplitude 
through the collimator response function. 

Upper limits were estimated on the flux from XTE J0052-723 in the observation 
preceeding the first detection and following the last detection. The power in the Lomb-Scargle periodogram at the pulse period
in these observations was converted into pulse amplitude and corrected for collimator response, giving values of 0.2 and 0.5 counts/PCU/s.
Because the pre-outburst observation contained other pulsars (see above) the upper limit was also estimated by alternative means. 
Simulated data consisting of white noise 
plus a periodic modulation were sampled onto the window function of the actual observation 
and analysed as described in Section~\ref{section:timing}.
The procedure was repeated, incrementally increasing the signal amplitude until positive 
detection occurred. Two different types of modulation were considered in order to investigate any effects 
of pulse profile on sensitivity. The upper limit for a sinewave plus white noise 
was $\sim$0.1 Counts PCU$^{-1}$s$^{-1}$ pulsed amplitude, rising to $\sim$0.3 Counts 
PCU$^{-1}$s$^{-1}$ for a template derived from the actual pulse profile seen in 
observation 6 (see Figure~\ref{fig:profiles}). 

\subsection{PCA X-ray Spectrum}
\label{section:xspec}
\begin{figure}
\begin{center}
\psfig{file=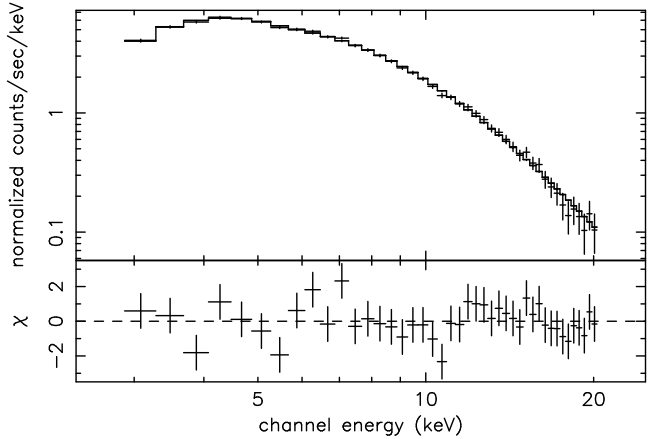,width=8.5cm,angle=-90}
\caption{PCA spectrum for observation 6. Details of the high energy cutoff powerlaw 
model are given in Section~\ref{section:xspec}.}
\label{fig:spectrum}
\end{center}
\end{figure}

The X-ray spectrum of XTE J0052-723 (Figure~\ref{fig:spectrum}) was obtained in 
observation 6 with the PCA pointing directly at the best-fit position.  
All 5 PCUs were on during the first 2.5 ksec of this observation, the spectrum was 
extracted from the Good Xenon data for PCUs 1-4, PCU0 was excluded 
(see Section~\ref{section:detection}). Background subtraction was performed with the L7 model 
for faint sources. Spectra were not extracted for the other pointed observations in which XTE J0052-723 was
$\sim$0.8\degr off axis.

Of the spectral models tested, the spectrum was well fitted by a power 
law with high energy cutoff ($\alpha$=1.77, $E_{cut}$=14 keV, $E_{fold}$=16 keV, 
$n_{H}$=3.6$\times$10$^{22}$ cm$^{-2}$) and by bremsstrahlung (kT=19.47 keV, $n_{H}$=2.1$\times$10$^{22}$ cm$^{-2}$) 
$\chi_{\nu}^{2}$(DOF) for the two models was 1.13(38) and 0.92(40) respectively. A simple absorbed power 
law model gave a significantly worse $\chi_{\nu}^{2}$ and proved a poor representation of the high energy 
end of the spectrum, demonstrating the the existence of a high energy cut-off above about 15 keV. 
A blackbody model was not able to fit the data. The suitability of the cutoff powerlaw model for 
X-ray pulsars has been discussed by e.g. White, Swank \& Holt \shortcite{white83}.
The $unabsorbed$ luminosity (cut-off power law) of XTE J0052-723 using D$_{SMC}$=63 kpc \cite{gro1} was found to 
be L$_{2-10keV}=$5.3$\times$10$^{37}$ erg/s, L$_{10-20keV}=$2.9$\times$10$^{37}$ erg/s. 
Assuming the fraction attributable to other sources (see Section~\ref{section:timing}) is around 12\% 
(0.9 counts in 7.52 for this observation) these luminosities scale down 
to 4.7 and 2.3 for a total $L_{X}$=7$\times$10$^{37}$erg/s. 
We note that the parameters obtained for the high energy cut-off model were obtained by starting the
fitting process at reasonable values as given by White, Swank \& Holt \shortcite{white83}. If instead no 
$a priori$ assumptions were made, and all parameters set to unity, the fit converged to a lower
$\chi_{\nu}^{2}$ but with a cut-off energy which is unusually low for an accreting pulsar 
($E_{cut}$=6.4 keV, $E_{fold}$=27 keV). The value of 6.4 keV is strongly suggestive of the presence of 
an iron K line which has been observed in many pulsars (Nagase \shortcite{Nag89}). This possibility was investigated by 
adding a gaussian component to the model. The line energy was set initially at 6.4 keV but was fitted as a free parameter. 
(Best fitting line energy 6.399$\pm$0.22 keV, width 1.4$\pm$0.8keV, EW=281 eV). 
This fit gave parameters consistent with our original 
result ($\alpha=1.77\pm0.08$,$E_{cut}=15\pm1$ keV, $E_{fold}=12\pm6$ keV,
$n_{H}=2.5\pm0.8\times10^{22} cm^{-2}$) and is a slight improvement
$\chi_{\nu}^{2}$(DOF)=0.75(35). The SMC background contribution mentioned above, 
may also contribute to peculiarities in the spectral fitting.

\section{Optical Observations} 
\subsection{Photometry}
\label{section:phot}
\begin{figure}
\begin{center}
\psfig{file=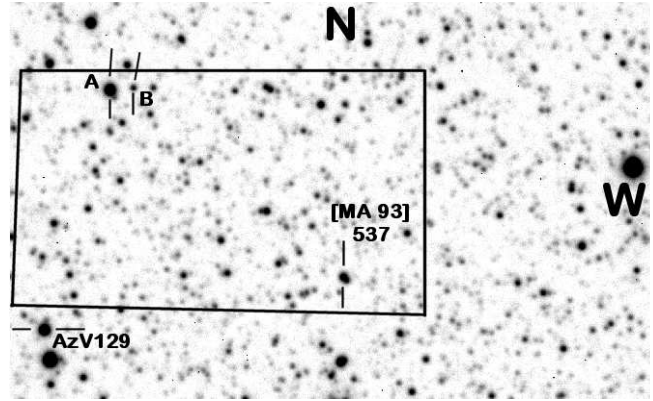,width=8.5cm,angle=-90}
\caption{CCD Johnson R image covering a 5\arcmin wide field. The rectangle is   
the XTE 99\% error box for the new pulsar. The optical counterpart is
probably [MA93]537.}  
\label{fig:fc}
\end{center}
\end{figure}

In order to try and identify the counterpart, optical observations
were made at SAAO with the 1m telescope on 16th January 2001 while the
X-ray pulsar was in outburst. The STE4 CCD chip was used which gives a
5$\times$5 arcminute field. The region of sky included the full area of
the XTE 99\% confidence region (Figure~\ref{fig:fc}), and it was imaged in
Johnson B,V,R and narrow-band H$\alpha$ filters.  Full field aperture
photometry was performed on the images and B and V instrumental
magnitudes were calibrated against a star in the field listed in the
TYCHO2 catalogue \cite{tycho2}. The chosen calibration star was TYC
9138-1734-1. The resulting B and V magnitudes for the stars indicated
are given in Table~\ref{tab:phot}.
\begin{figure}
\begin{center}
\psfig{file=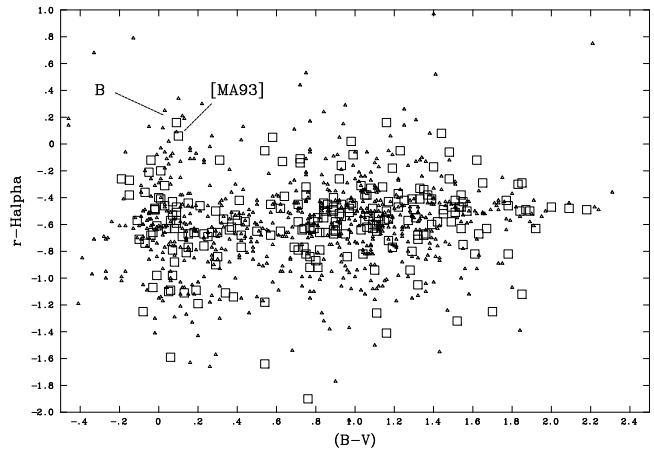,width=8.5cm,angle=-90}
\caption{Optical colour-colour diagram (see Section~\ref{section:phot}) 
 for the field shown in Figure~\ref{fig:fc}. The two objects within the error
 region that have the correct colours for a B0-B2 star and 
 the strongest H$\alpha$ emission are marked. 
 $\sq$= stars in the RXTE error box, $\bigtriangleup$= other stars in the field}
\label{fig:cmd}
\end{center}
\end{figure}  
\begin{table}
 \caption{Optical photometry of candidates}
 \label{tab:phot}
 \begin{tabular}{lccccc} 
 Object    & Bmag     &  Vmag  & Jmag$^{\star}$  & Hmag$^{\star}$  & Kmag$^{\star}$ \\
\hline
 A         & 15.8      & 14.3  & 	&	 &			\\
 B         & 17.4      & 17.3  &	&	 &         		\\
 $[MA93]$ 537 & 15.9   & 15.8  & 16.46  & 15.65  & 15.7    \\
 AzV129    & 14.3      & 14.3  & 14.46  & 14.59  & 14.6    \\
\end{tabular}
\medskip
$^{\star}$Infrared magnitudes from NASA/IRSA catalogue.
\end{table}

The resulting magnitudes were then used to plot the colour-colour
diagram: B-V versus R-H$\alpha$, shown in Figure~\ref{fig:cmd}. The
calibrated (B-V) colour axis compares favourably with the colour-magnitude
diagram for the SMC of Udalski et al. \shortcite{udalski98}.
These plots were used to identify candidates for an optical
counterpart to the X-ray pulsar. The B-V colour provides a measure of
effective stellar temperature, with B0-B2 objects (the typical
counterparts to these systems) lying in the (B-V) = 0$\pm$0.2 region.
The R-H$\alpha$ value estimates the strength of the H$\alpha$ emission
line relative to the surrounding (red) continuum.  For hot blue stars
lying to the left of this plot, a relatively large R-H$\alpha$ cannot
be consistent with the stellar emission and implies the presence of a
significant mass of cooler material radiating at longer
wavelengths. This situation provides the photometric classification of
a Be star \cite{jaschek}.
 
The R band image of the field obtained at SAAO is shown in
Figure~\ref{fig:fc}. The XTE 99\% confidence error box is outlined and
Be stars identified by the above procedure are indicated. Each candidate
is cross-referenced to its position on the 2-colour
diagram (Figure~\ref{fig:cmd}).  Two candidate Be stars appear in the
XTE error box, the brighter of the two is a known emission-line star
[MA93]537 \cite{MA93}. The one labelled Object B has a slightly more extreme
R-H$\alpha$ but is much fainter.

\subsection{Spectroscopy}
\label{section:spec}
\begin{figure} 
\begin{center}
\psfig{file=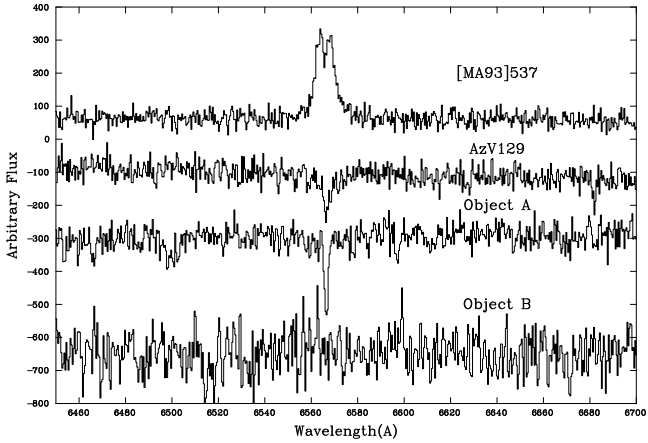,width=8.5cm,angle=-90}
\caption{H$\alpha$ spectra of the stars [MA93]537, AzV129, Object A, 
 and Object B.} 
\label{fig:spectra}
\end{center}  
\end{figure}

H$\alpha$ spectra were taken on November 7th 2001 with the SAAO 1.9m,
for four objects in the field. These were the two objects identified 
from the photometry, plus AzV129 (a 14th mag B1 star \cite{AzV}) and 
object A which was in the slit with object B. Only [MA93]537 shows 
H$\alpha$ in emission, with a strong double peaked line-profile shown in
Figure~\ref{fig:spectra}.  Such a double-peaked line implies that
H$\alpha$ emission comes from a circumstellar disk \cite{slet92}, the
disk providing the reservoir of fuel to power the X-ray pulsar.
The equivalent width was determined to be -43.3 $\pm$ 0.7\AA and the
line was centred at 6566\AA. The separation velocity of the peaks was
found to be 200 $\pm$ 15 km/s.

\section[]{Discussion} 
The position of the H$\alpha$ peak in [MA93]537 corresponds to a red
shift of 137 km/s. The agreement of this value with the general
recessional (red-shift) value for the SMC of 148 km/s confirms its membership of
that body. The observed V magnitude of this object corresponds to a
B0V-B1V star in the SMC, though the (B-V) colours are slightly too
blue. However, close inspection of the image for [MA93]537 reveals it
to be a double star with one object approximately 3-4 times brighter
than the other in the R band. It is quite possible that the blend of the
fainter star with the brighter one is affecting the colours in the
manner observed. Notwithstanding this small effect, there can be
little doubt that [MA93]537 must be the prime candidate for the optical
counterpart to the X-ray source. Table~\ref{tab:phot} also lists archival
infrared magnitudes which show a rising flux into the IR, it should be 
noted that these values are not from the same epoch as the SAAO photometry.  

The maximum $L_{X}$ attained by XTE J0052-723 reached $\sim$10$^{38}$erg/s
as inferred from its luminosity in observation 6 and its peak 2-10 keV
flux (observation 2).  The outburst appears to have brightened rapidly from
$<$8$\times$ 10$^{36}$erg/s (upper limit) to $\leq$10$^{38}$erg/s within $\sim$6 
days, and then declined slowly over the next month.

Using the relation of Bohlin et al. \shortcite{bohlin78}
the measured $N_{H}$ implies E(B-V)=3.55, the apparent colours of the
proposed counterpart (Table~\ref{tab:phot}) do not support this degree
of reddening, which if in front of the star would imply B-V $<$ -3.
Thus the large $N_{H}$ must be due to local material around the neutron star.

\begin{figure}
\begin{center}
\psfig{file=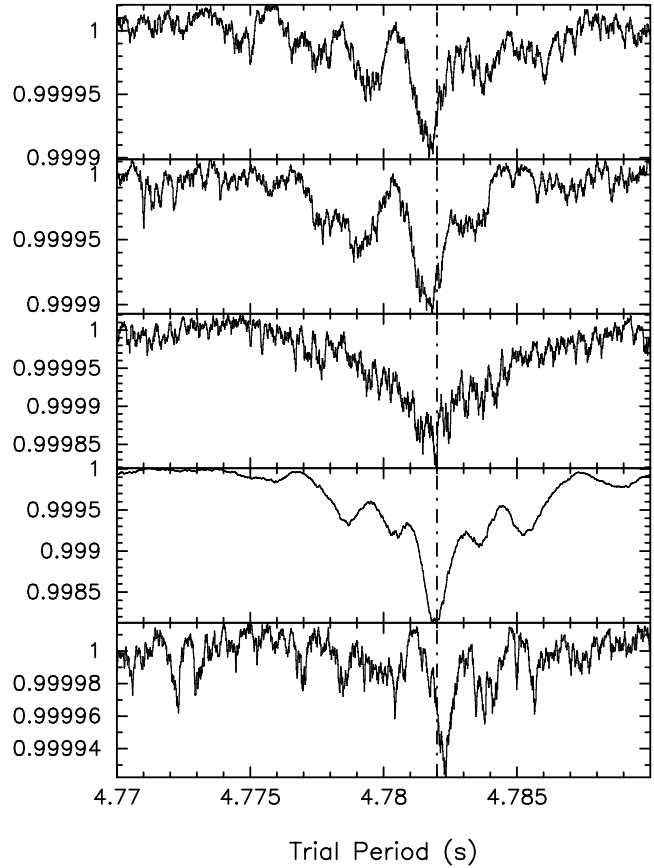,width=8.5cm}
\caption{PDM periodograms centered on the 4.782s feature. A drift in pulse 
period is observed over the 28 days of the outburst. From top down, 
observations 1,2,4,6,7}
\label{fig:pdm}
\end{center} 
\end{figure}

The pulse profile (Figure~\ref{fig:profiles}) is double peaked and rather asymmetric at all times. 
Timing analysis shows evidence for changes in pulse period between the observations.
This is revealed in Figure~\ref{fig:pdm} as an apparently continuous shift in the
position of the 4.782s feature in the PDM periodogram during the 28 days spanned by our 
observations. The total period change is 6.4$\times$10$^{-4}$s giving a mean period
derivative ($\dot{P})$ of 2.65$\times$10$^{-10}$s/s.The direction of the effect is 
an apparent lengthening of the pulse period.   
Since the distance to the SMC is known to within 10\% we can estimate the likely contributions to
pulse period variation, due to accretion torques and binary motion.
The maximum possible spin derivative given the observed luminosity, was calculated 
using the standard model for accretion torque as described by, for example, Rappaport \& Joss
\shortcite{rapp77}. 

\[
L_x \ge \frac{I\dot {\nu }}{R_x }\left( {16\pi ^4GM_{x} \nu }
\right)^{\raise0.7ex\hbox{$1$} \!\mathord{\left/ {\vphantom {1
3}}\right.\kern-\nulldelimiterspace}\!\lower0.7ex\hbox{$3$}}
\]

Re-arranging for $\dot{\nu }$ and solving for a 1.4M$_{\sun}$ neutron star of radius 
15 km radiating 10$^{38}$ erg/s and spinning once every 4.782 sec, gives a maximum 
spin-up rate of 1.52$\times$10$^{-11}$ Hz/s. Since $\nu$/$\dot{\nu}$= -$P/\dot{P}$ 
we have $\dot{P}_{max}$=3.49$\times$10$^{-10}$s/s. 
This is similar to the observed $\dot{P}$ but of the opposite sign.
We note that spin-down as well as spin-up has been seen for a number of X-ray pulsars
e.g. Bildsten \shortcite{bildsten98}, and that the spin-down rates may be similar to the
spin-up rates.
  
If XTE J0052-723 is similar to other Be HMXBs then its orbital period 
may be expected from the $P_{Pulse}/P_{Orbit}$ diagram \cite{corbet86} 
to be $\sim$20 to 40 days, comparable with the duration of this outburst. 

To estimate the possible contribution of orbital doppler shifts to the observed pulse
period changes we made a simple estimate of the neutron star's orbital velocity and
compared this to the projected velocity implied by the observed change in pulse period.  
We assume a $10M_{\sun}$ primary, orbital period of 30 days, circular orbit, negligible mass 
secondary. The orbital radius for such a system is 6.1$\times$10$^{7}$ km, and the orbital velocity
is 147 km/s. If the correct optical counterpart is the Be star we have identified, then the inclination
of the system must be fairly high to generate the double-peaked spectral line profile. 
Taking the maximum and minimum measured values for the pulse period as given in Table~\ref{tab:periods} 
implies a velocity shift of 38 km/s. 

We therefore conclude that the observed variations in the pulse period may have
contributions from both orbital doppler shifts and accretion torques. In order to measure
the contributions of both of these effects, and to determine the orbital parameters of the
system, more extensive pulse timing measurements are required. 

\section{Conclusions}
XTE J0052-723 has been identified as a transient X-ray pulsar in
the SMC. It was observed in X-ray outburst from 27/12/00 - 16/01/01. Non-detections 
before and after constrain the  duration of the outburst to be 36$\pm$5days.
The peak X-ray luminosity reached $\sim$10$^{38}$ergs/s. An X-ray spectrum obtained  
during the decline of the outburst was equally well fitted by a 
cut-off powerlaw ($\alpha$=1.7, $E_{cut}$=15 keV) or bremsstrahlung model ($kT$=19.47 keV), 
both requiring a large amount of absorbing gas ($N_{H}>2\times10^{22}$cm$^{-2}$) around the neutron star.
A pulse-period derivative was detected, with an amplitude consistent with that expected for binary motion. 
The X-ray source position was localised by performing slews. 
Optical photometry was used to search for an optical counterpart during the X-ray outburst. 
Spectroscopy of selected candidates has identified the probable counterpart, which 
is a BOV-B1V SMC member, exhibiting a strong, double peaked H$\alpha$ emission
line (EW=-43.3\AA, V$_{Separation}$=200$\pm$15 km/s). Archival infrared magnitudes
indicate an IR excess as expected for a Be star.  

This discovery further increases the number of known HMXBs in this remarkable galaxy.
It is becoming increasingly clear that the SMC contains an unusually high abundance of
massive stars and the domination of the HMXB population by Be/X-ray binaries is
intriguing. Our $RXTE$ monitoring program continues to follow the outbursts of the 
SMC's X-ray pulsars and results on the long term behaviour of a large fraction of that
population will be published in the near future.

 \section*{Acknowledgments}
SGTL is in receipt of a PPARC studentship. This work has made use of
the SIMBAD database, operated at CDS, Strasbourg, France. The optical 
observations were made at the South African Astronomical Observatory.

\label{lastpage}

\begin{thebibliography}{}

\bibitem[\protect\citename{Azzopardi \& Vigneau }1978]{AzV}
 Azzopardi M., Vigneau J., 1978, A\&AS, 50, 291
\bibitem[\protect\citename{Bildsten }1998]{bildsten98}
 Bildsten L., 1998, "The Angular Momentum of Accreting Neutron Stars",
 in "Accretion Processes in Astrophysical Systems: Some Like it Hot!",
 AIP Conference Proceedings 431, ed. S.S. Holt \& T.R. Kallman, pp. 299-308
\bibitem[\protect\citename{Bildsten et al. }1997]{bildsten97}
 Bildsten L., Chakrabarty D., Chiu J., Finger M. H., Koh T., 
 Nelson R. W., Prince T. A., Rubin B. C., Scott D. M., Stollberg M., 
 Vaughan B. A., Wilson C. A., Wilson R. B. 1997, ApJS, 113, 367 
\bibitem[\protect\citename{Bohlin et al. }1978]{bohlin78}
 Bohlin R. C., Savage B. D., Drake J. F., 1978, ApJ, 224, 132
\bibitem[\protect\citename{Coe }2000]{coe00}
 Coe M. J., 2000, in "The Be Phenomenon in Early-Type Stars",
 IAU Colloquium 175, ASP Conference Proceedings, Vol. 214, eds M. A. Smith
 and H. F. Henrichs. Astronomical Society of the Pacific, p.656
\bibitem[\protect\citename{Corbet }1986]{corbet86}
 Corbet, R. H. D., 1986, MNRAS, 220, 1047
\bibitem[\protect\citename{Corbet et al. }2001]{corbet2001}
 Corbet R. H. D., Marshall F. E., Markwardt C. B., 2001, IAUC, 7562
\bibitem[\protect\citename{Gardiner \& Noguchi }1996]{GN96}
 Gardiner L. T., Noguchi M., 1996, MNRAS, 278, 191
\bibitem[\protect\citename{Gardiner }1999]{G99}
 Gardiner L. T., 1999, in "New Views of the Magellanic Clouds",
 IAU Symposium, Vol. 190, Y.-H. Chu, N.B. Suntzeff,
 J.E. Hesser \& D.A. Bohlender, eds.
\bibitem[\protect\citename{Groenewegen }2000]{gro1} 
 Groenewegen M. A. T., 2000, A\&A, 363, 901
\bibitem[\protect\citename{Haberl \& Sasaki }2000]{hs00}
 Haberl F., Sasaki M., 2000, A\&A, 359, 573
\bibitem[\protect\citename{Hog et al. }2000]{tycho2}
 Hog E., Fabricius C., Makarov V. V., Urban S., Corbin T., 
 Wycoff G., Bastian U., Schwekendiek P., Wicenec A., 2000, A\&A, 355, L27
\bibitem[\protect\citename{Jahoda et al. }1996]{jahoda96}
 Jahoda K., Swank J. H., Stark M. J., Strohmayer T., Zhang W., Morgan E. H.,
 1996, Proc.SPIE,2808,59
\bibitem[\protect\citename{Jascheck \& Jaschek }1990]{jaschek}
 Jaschek C., Jaschek M., 1990, The Classification of Stars, 
 Cambridge University Press 
\bibitem[\protect\citename{Laycock et al. }2002]{lay01}
 Laycock S., Corbet R. H. D., Perrodin D., Coe M. J., Marshall F. E., Markwardt 
 C., 2002, A\&A, in press, astro-ph/0201468
\bibitem[\protect\citename{Li \& Thronson }1999]{LT99}
 Li P. S., Thronson H. A., 1999, in "New Views of the Magellanic Clouds", 
 IAU Symposium, Vol. 190, Y.-H. Chu, N.B. Suntzeff, 
 J.E. Hesser \& D.A. Bohlender, eds.
\bibitem[\protect\citename{Maragoudaki et al. }2001]{M01}
 Maragoudaki F., Kontizas M. Morgan D. H., Kontizas E., Dapergolas A., Livanou E.,
 2001, A\&A, 379, 864
\bibitem[\protect\citename{Meyssonnier \& Azzopardi }1993]{MA93}
 Meyssonnier N., \& Azzopardi, M., 1993, A\&AS, 102, 451
\bibitem[\protect\citename{Nagase }1989]{Nag89}
 Nagase F., 1989, PASJ, 41, 1
\bibitem[\protect\citename{Negueruela }1998]{neg98}  
 Negueruela I., 1998, A\&A 338, 505 
\bibitem[\protect\citename{Porter }1996]{porter96}
 Porter J., 1996, MNRAS, 280, L31
\bibitem[\protect\citename{Press \& Rybicki }1989]{press89}
 Press W. H., Rybicki G. B., 1989, ApJ, 338, 277 
\bibitem[\protect\citename{Press et al. }1993]{press93}
 Press W. H., Teukolsky S. A., Vetterling W. T., Flannery B. P., 
 Numerical Recipies in Fortran 2nd Ed., 1993, Cambridge University Press
\bibitem[\protect\citename{Rappaport \& Joss }1977]{rapp77}
 Rappaport, S. \& Joss, P. C. 1977, Nature, 266, 683 
\bibitem[\protect\citename{Scargle }1982]{sca82}
 Scargle J. D., 1982, ApJ, 263, 835
\bibitem[\protect\citename{Singh et al. }1995]{singh95}
 Singh K. P., Barrett P., Schlegel E., White N. E., 1995, IAUC 6195
\bibitem[\protect\citename{Slettebak et al. }1992]{slet92}
 Slettebak A., Collins G. W., Truax R., 1992, ApJS, 81,335
\bibitem[\protect\citename{Stanimirovic et al. }1997]{S99}
 Stanimirovic S., Staveley-Smith L., Dickey J. M., Sault R. J., Snowden S. L., 
 1999, MNRAS, 302, 417
\bibitem[\protect\citename{Stellingwerf }1978]{stel78}
 Stellingwerf R. F., 1978, ApJ, 224, 953
\bibitem[\protect\citename{Udalski et al. }1998]{udalski98}
 Udalski A., Szymanski M., Kubiak M., Pietrzynski G., Wozniak P., Zebrun K., 
 1998, Acta Astron., 48, 147 
\bibitem[\protect\citename{van den Heuvel }1983]{vdh83}
 van den Heuvel E. P. J., 1983, In: Accretion Driven Stellar X-ray Sources, 
 Lewin W. H. G. \& van den Heuvel E. P. J., eds. 
\bibitem[\protect\citename{White et al. }1983]{white83}
 White, N. E, Swank, J. H., Holt, S. S., 1983, 
 ApJ, 263, 277
\bibitem[\protect\citename{Yokogawa et al. }2000]{yok00}
 Yokogawa J., Torii K., Imanishi K., Koyama K., 2000, PASJ, 52, L37
\bibitem[\protect\citename{Yoshizawa \& Noguchi }1999]{YN99}
 Yoshizawa A. M., Noguchi M., Galaxy Interactions at Low and High Redshift, 
 IAU Symposium 186, Edited by J. E. Barnes \& D. B. Sanders, 1999, p.60 
\bibitem[\protect\citename{Zaritsky et al. }2000]{Z00}
 Zaritsky D., Harris J., Grebel E. K., Thompson I. B., 2000, ApJ, 534, L53
 
\end{thebibliography}
\end{document}